\documentstyle[aps,12pt,epsfig,manuscript]{revtex}
\global\firstfigfalse
\begin{document}
\preprint{INJE-TP-99-5}
\def\overlay#1#2{\setbox0=\hbox{#1}\setbox1=\hbox to \wd0{\hss #2\hss}#1%
\hskip -2\wd0\copy1}

\title{Nonpropagation of tachyon on the BTZ black hole 
in type 0B string theory}

\author{ H.W. Lee$^1$, Y.S. Myung$^1$ and  Jin Young Kim$^2$ }
\address{$^1$Department of Physics, Inje University, Kimhae 621-749, Korea\\
$^2$ Department of Physics, Kunsan National University, Kunsan 573-701, Korea}

\maketitle

\begin{abstract}
We obtain the BTZ black hole (AdS$_3 \times$S$^3$) as a 
non-dilatonic solution from type 0B string theory.
Analyzing the s-wave perturbation around this black hole background, 
we show that the tachyon is not a propagating mode.
\end{abstract}

\newpage
\section{Introduction}
\label{introduction}
Recently type 0 string theories attract much interest
in the study of non-supersymmetric gauge theories
\cite{Pol99IJMPA645,Kle99NPB155,Min99JHEP01020,Dix86NPB93,Fer9811208,Kle99JHEP03015}.
Type 0 string theories can be obtained from the worldsheet of 
type II string theories by performing a non-chiral GSO projection
\cite{Dix86NPB93}.
The resulting theories have world sheet supersymmetry but no 
spacetime supersymmetry.
The crucial difference of type 0 theories with type II theories 
is to have the doubling of Ramond-Ramond(RR) fields and the tachyon. 

One of the simplest way to see the role of tachyon in type 0 theories is
to consider the intersecting D$p$-branes.
The D$p_{\pm}$-brane bound states 
can be intersected according to the same rule of the type II theories.
The D$5_{\pm}$-D$1_{\pm}$ brane black hole was constructed 
to show that the corresponding near-horizon geometry is 
AdS$_3\times$S$^3\times$T$^4$ and it has asymptotically flat 
space at infinity\cite{Cos9903128}.
This corresponds to the dilatonic solution.
It is shown that the tachyon field can be stabilized 
only in the near-horizon of 
AdS$_3 \times$S$^3 \times$T$^4$. 
In our previous work\cite{Lee9907024}, we studied extensively the stability by 
analyzing the potentials surrounding the D5$_\pm$-D1$_\pm$ 
brane black hole.
In this paper, we find the BTZ black hole as a non-dilatonic solution 
from type 0B string theory\cite{Sac9907201}.
This is a globally AdS$_3 \times$S$^3$ solution, which means that it 
has asymptotically AdS$_3$ spacetime\cite{Lee9805050}.
This is a crucial point that contrasts to 
the D5$_\pm$-D1$_\pm$ brane black hole.
Here we wish to study whether the tachyon can propagate or not on global 
AdS$_3 \times$S$^3$.

The organization of the paper is as follows.
In section \ref{blackhole}, we obtain the BTZ
black hole in type 0B string theory.
We set up the s-wave perturbation for all fields around this black hole 
background in Sec. \ref{perturbation}.
Here we choose the harmonic gauge for graviton and use all linearized equations 
including two Bianchi identities 
to decouple ($\phi,t$) from the remaining fields.
Finally we discuss our results in Sec. \ref{discussion}.

\section{BTZ Black Hole}
\label{blackhole}
We start with the appropriate truncation of the type 0B string theory 
in the string frame\cite{Kle99NPB155,Cos9903128}
\begin{eqnarray}
S_{10} &=& { 1\over \kappa_{10}^2}
\int d^{10}x \sqrt{-g} 
\left [ 
e^{-2 \Phi} \left \{ R + 4 (\nabla \Phi)^2 
- {1 \over 4} \left ( \nabla T \right )^2 - {m^2 \over 4} T^2
\right \} \right .
\nonumber \\
&&
\qquad\qquad\qquad\qquad\qquad\qquad\qquad\qquad
\left .
- {1 \over 12}  
\left \{
f_+(T) F_{+3}^2 + f_-(T) F_{-3}^2
\right \}
\right ]
\label{string-action}
\end{eqnarray}
where $f_{\pm}(T) = 1 \pm T + {1 \over 2} T^2$, 
$m^2 = -2 /\alpha', \kappa_{10}^2 = (2 \pi)^6 \pi g^2 \alpha'^4$
and $F_{\pm3}$ are the Ramond-Ramond(RR) three-forms.
Comparing with the results of type IIB theory
\cite{Cal97NPB65,Lee98PRD104006}, the new ingredients are the 
tachyon($T$) and the doubling of the RR fields($F_{\pm3}$).
Here $\alpha'=1$ and 
$g$ is ten-dimensional string coupling constant.  
We wish to follow the MTW conventions\cite{Mis73}.

The equations of motion for action (\ref{string-action}) are given by
\begin{eqnarray}
&&R_{MN} 
+ { 1 \over 4} g_{MN} \nabla^2 \Phi
+2 \nabla_M \nabla_N \Phi
- {1 \over 2} g_{MN} \left ( \nabla \Phi \right )^2
\nonumber \\
&&\qquad\qquad
- { 1 \over 4} \nabla_M T \nabla_N T
- {m^2 \over 32} g_{MN} T^2
+{ 1 \over 48} g_{MN} e^{2 \Phi} 
\left \{
f_+(T) F_{+3}^2 + f_-(T) F_{-3}^2
\right \}
\nonumber \\
&&\qquad\qquad
- { 1 \over 4} e^{ 2 \Phi}
\left \{
f_+(T) F_{+3MPQ} F_{+3N}^{~~~~PQ}
+f_-(T) F_{-3MPQ} F_{-3N}^{~~~~PQ}
\right \}
= 0,
\label{eq-ricci} \\
&& 
R + 4 \nabla^2 \Phi - 4 \left ( \nabla \Phi \right )^2 
- { 1 \over 4} \left ( \nabla T \right )^2 - { m^2 \over 4} T^2 
= 0, \label{eq-Phi}\\
&&
\nabla_M 
\left \{
f_+(T) F_{+3}^{~~MNP}
\right \}
=0,
   \label{eq-fp} \\
&&
\nabla_M 
\left \{
f_-(T) F_{-3}^{~~MNP}
\right \}
=0,
   \label{eq-fm} \\
&&
\nabla^2 T -2 \nabla \Phi \nabla T - m^2 T 
-{ 1 \over 6} e^{2\Phi}
\left \{
f_+'(T) F_{+3}^2 + f_-'(T) F_{-3}^2
\right \}
=0,
\label{eq-t}
\end{eqnarray}
where the prime($'$) denotes the differentiation with respect to its argument.
From Eqs.(\ref{eq-ricci}) and (\ref{eq-Phi}), one can rewrite the new 
dilaton equation as
\begin{equation}
\nabla^2 \Phi - 2 \left ( \nabla \Phi \right )^2
- { m^2 \over 8} T^2
- { 1 \over 12} e^{2 \Phi}
\left \{
f_+(T) F_{+3}^2 + f_-(T) F_{-3}^2
\right \}
= 0. \label{eq-dilaton}
\end{equation}
In addition, we need the remaining Maxwell equations as two 
Bianchi identities\cite{Lee98PRD104006} 
\begin{equation}
\partial_{[M}F_{\pm3NPQ]}=0.
\label{bianchi}
\end{equation}

Here we are interested in non-dilatonic solution. 
We consider mainly the six-dimensional part by parametrizing 
$g_{10} = e^{\phi_6} g_6 + e^{2 \chi} dx_i dx^i$, where 
$\phi_6 = \Phi - 2 \chi$.
Note the difference between $\Phi$ and $\phi_6$.
The former is the 10D dilaton and the latter is the 6D dilaton.
Hereafter we set $\phi_6$ to be zero, which means that 
\begin{equation}
\Phi = 2 \chi.
\label{Phi-chi}
\end{equation}
The black hole solution can be obtained by setting
\begin{equation} 
e^{2 \bar \Phi} = g^2, ~ \bar T = 0, ~
{\bar F}_{\pm 3}^2 = 0, ~ \bar R =0 .
\label{bck-sol}
\end{equation}
In detail one finds
\begin{eqnarray}
{\bar F}_{\pm3} &=&
{ \sqrt{2} r_5^2 \over g} \epsilon_3
+ \sqrt{2} r_1^2 g e^{-2 \bar \Phi} *_6 \epsilon_3,
\label{back-F}\\
{\bar R}_{\mu\nu} &=&
- {2 \over R^2} {\bar g}_{\mu\nu}^{\rm BTZ}, \quad
{\bar R}_{mn} =
{2 \over R^2} {\bar g}_{mn}^{{\rm S}_3},
\label{bck-R}
\end{eqnarray}
with $R^2 = r_1 r_5$ and $*_6 \epsilon_3$ is the six-dimensional Hodge dual 
of $\epsilon_3$.
In type 0B string theory, the above solution was first suggested 
in \cite{Cos9903128}.
The ten-dimensional indices $M, N, P, \cdots$ are split into 
$\mu,\nu,\rho, \cdots$ for the BTZ black hole
$(t, \rho, \varphi)$ and $m,n,p, \cdots$ for 
S$^3(\theta_1, \theta_2, \theta_3)$.
The background spacetime (AdS$_3 \times $S$^3 \times$T$^4$) is 
given by  
\begin{equation}
ds_{10}^2 = 
ds_{\rm BTZ}^2 + R^2 d \Omega_3^2 + dx_i^2
\label{metric}
\end{equation}
where the BTZ black hole spacetime is given by\cite{Ban92PRL1849}
\begin{equation}
ds_{\rm BTZ}^2 = - {{(\rho^2 - \rho_+^2)(\rho^2 -\rho_-^2)} \over
   {\rho^2 R^2 }} dt^2 + \rho^2 ( d \varphi - {J \over 2 \rho^2} dt )^2 +
  { \rho^2 R^2 \over {(\rho^2 - \rho_+^2)(\rho^2 -\rho_-^2)}} d \rho^2.
\label{BTZ-metric3}
\end{equation}
Here $M = (\rho_+^2 + \rho_-^2)/R^2, J=2 \rho_+ \rho_- /R$ are the mass and
angular momentum of the BTZ black hole.
In this case, the relevant
thermodynamic quantities(Hawking temperature, area of horizon,
angular velocity at horizon, left/right temperatures) are given by
\begin{eqnarray}
T_H^{\rm BTZ} &=& {\rho_+^2-\rho_-^2 \over 2 \pi R^2 \rho_+} ,
\nonumber \\
{\cal A}_H^{\rm BTZ} &=& 2 \pi \rho_+,~~ \Omega_H = { J \over 2 \rho_+^2},
\\ \label{temperature}
{1 \over T_{L/R}^{\rm BTZ}} &=& {1 \over T_H^{\rm BTZ}}
    \left ( 1 \pm {\rho_+ \over \rho_-} \right ) .
\nonumber
\end{eqnarray}

\section{Perturbation Analysis}
\label{perturbation}
For simplicity,  
we consider the s-wave perturbation around the 6D background of 
AdS$_3\times$S$^3$ as in \cite{Lee98PRD104006}
\begin{eqnarray}
&&F_{\pm3\mu\nu\rho} = 
{\bar F}_{\pm3\mu\nu\rho} + {\cal F}_{\pm3\mu\nu\rho} = 
{\bar F}_{\pm3\mu\nu\rho} \left \{ 1 + 
{\cal F}_{\pm}(t,\rho,\varphi,\theta_1, \theta_2, \theta_3)\right \},
\label{ptrFBTZ} \\
&&F_{\pm3mnp} = 
{\bar F}_{\pm3mnp} + {\cal F}_{\pm3mnp} = 
{\bar F}_{\pm3mnp} \left \{1 + 
{\cal F}_{\pm}^\theta(t,\rho,\varphi,\theta_1, \theta_2, \theta_3)\right \},
\label{ptrFS3} \\
&&\Phi = \bar \Phi + \phi(t,\rho,\varphi,\theta_1, \theta_2, \theta_3),  
\label{ptr-Phi}  \\  
&&g_{MN} = \bar g_{MN} + h_{MN}(t,\rho,\varphi,\theta_1, \theta_2, \theta_3),
\label{ptrg} \\
&& T = 0 + t(t,\rho,\varphi,\theta_1, \theta_2, \theta_3).
\label{ptr-t}
\end{eqnarray}
We remind the reader that the forms of perturbation should be at 
least taken to preserve the background symmetry of AdS$_3 \times$S$^3$.
In this sense we choose a diagonal form for
${\cal F}_{\pm3MNP}$.
Here $h_{MN}$ is given by the block diagonal 
form\cite{Lee9805050,Gre94NPB399}
\begin{equation}
h_{MN} = \left [
\begin{array}{ccccccc}
h_{tt} & h_{t \rho} & h_{t \varphi} & &&&\\
h_{\rho t} & h_{\rho \rho} & h_{\rho \varphi} & &0 &&0 \\
h_{\varphi t} & h_{\varphi \rho} & h_{\varphi \varphi} & &&&\\
&&&h_{\theta_1 \theta_1} & h_{\theta_1 \theta_2 } & h_{\theta_1 \theta_3} &\\
&0 &&h_{\theta_2 \theta_1} & h_{\theta_2 \theta_2 } & h_{\theta_2 \theta_3} &0\\
&&&h_{\theta_3 \theta_1} & h_{\theta_3 \theta_2 } & h_{\theta_3 \theta_3} &\\
&0&&&0 & &h_{ij}
\end{array}
\right ] .
\label{hMN}
\end{equation}
It is noted that (\ref{hMN}) is chosen to preserve the background symmetry 
of AdS$_3 \times$S$^3 \times$T$^4$.
This seems to be simple but it is sufficient for our s-wave calculation
with $l=0$.
Here $l$ is given by the relation as 
$\bar \nabla^2_{{\rm S}^3} \phi = -l(l+2) \phi$.
This is so because in s-wave calculation the graviton $h_{MN}$ are not 
propagating modes except $h_{ij}$.

One has to linearize (\ref{eq-Phi}), (\ref{eq-dilaton})  and 
(\ref{eq-t}) in order to obtain 
the equations governing the perturbations as
\begin{eqnarray}
&&\bar \nabla^2 (8\phi-h)
+ {4 \over R^2 } ( h^t_{~t} + h^{\rho}_{~\rho} + h^{\varphi}_{~\varphi}
 -  h^{\theta_i}_{~\theta_i})=0,
\label{eq-Phi-ptr} \\
&&\bar \nabla^2 \phi
+ {2  \over R^2 } \left [ ({\cal F}_+ +{\cal F}_- 
-{\cal F}_+^\theta -{\cal F}_-^\theta ) -
( h^t_{~t} + h^{\rho}_{~\rho} + h^{\varphi}_{~\varphi}
 -  h^{\theta_i}_{~\theta_i}) \right ]=0.
\label{eq-dilaton-ptr} \\
&&
\left ( \bar \nabla^2 - m^2 \right ) t
+ {4 \over R^2}
\left (
{\cal F}_+ - {\cal F}_- - {\cal F}_+^\theta + {\cal F}_-^\theta
\right ) =0,
\label{eq-t-ptr}
\end{eqnarray}
where $h^{\theta_i}_{~\theta_i} =
h^{\theta_1}_{~\theta_1} +  h^{\theta_2}_{~\theta_2} +
 h^{\theta_3}_{~\theta_3}$.
Now we attempt to disentangle the mixing between 
($\phi, t$)
and other fields by using both the harmonic gauge 
($\bar \nabla_M \hat h^{MP} = 0$, 
$\hat h^{MP} = h^{MP} - { h \over 2} \bar g^{MP}$, 
$h = h^Q_{~Q}$)
and Kalb-Ramond equations from (\ref{eq-fp}) and (\ref{eq-fm})
\begin{eqnarray}
&&\bar \nabla_M{\cal F}_{\pm3}^{MNP}
 - (\bar \nabla_M h^N_{~Q}) \bar F_{\pm3}^{MQP}
 + (\bar \nabla_M h^P_{~Q}) \bar F_{\pm3}^{MQN}
\nonumber \\
&&~~~~~~~~~
 - (\bar \nabla_M \hat h^M_{~Q}) \bar F_{\pm3}^{QNP}
 - h^M_{~Q} (\bar \nabla_M  \bar F_{\pm3}^{QNP})
 \pm (\partial_M t) \bar F_{\pm3}^{MNP}
=0,
\label{Kalb-Ramond}
\end{eqnarray}
When $N=t, P=\varphi$, solving Eq.(\ref{Kalb-Ramond}) leads to
\begin{equation}
\partial_\rho \left ( {\cal F}_\pm \pm t - h^t_{~t} - h^\varphi_{~\varphi} 
\right ) + \partial_t h^t_{~\rho} + \partial_\varphi h_\rho^{~\varphi} = 0.
\label{rel-tphi}
\end{equation}
Using the harmonic gauge, the last two terms turn out to be 
$\partial_\rho ( - h^\rho_{~\rho} + {1 \over 2} h)$.
Then (\ref{rel-tphi}) takes the form as
\begin{eqnarray}
2 \left ( {\cal F}_{\pm}  \pm t \right ) - 
\left ( h_{~t}^t + h_{~\rho}^\rho + h_{~\varphi}^\varphi \right ) 
+ h^{\theta_i}_{~\theta_i}  + h_{~i}^i &=& 0.
\label{rel-F} 
\end{eqnarray}
The remaining choices for $N, P$ lead to the same relation as in 
(\ref{rel-F}).
For $N=\theta_2, P=\theta_3$, one obtains the relation
\begin{eqnarray}
2 \left ( {\cal F}_{\pm}^\theta  \pm t \right ) + 
\left ( h_{~t}^t + h_{~\rho}^\rho + h_{~\varphi}^\varphi \right ) 
- h^{\theta_i}_{~\theta_i}  + h_{~i}^i &=& 0.
\label{rel-Ftheta}
\end{eqnarray}
From the Bianchi identities (\ref{bianchi}) one has
\begin{eqnarray}
&&\partial_{\theta_1} {\cal F}_\pm = \partial_{\theta_2} {\cal F}_\pm =
      \partial_{\theta_3} {\cal F}_\pm=0, 
\label{Bianchi-F} \\
&&\partial_t {\cal F}_\pm^\theta = \partial_\rho {\cal F}_\pm^\theta =
      \partial_\varphi {\cal F}_\pm^\theta=0. 
\label{Bianchi-Ftheta}
\end{eqnarray}
This implies that
${\cal F}_\pm= {\cal F}_\pm(t,\rho,\varphi)$ are
dynamical fields,
${\cal F}_\pm^\theta={\cal F}_\pm^\theta(\theta_1, \theta_2, \theta_3)$
are non-dynamical fields.
Hence we choose ${\cal F}_\pm^\theta=0$.
Then from (\ref{rel-Ftheta}) one obtains an important result as 
\begin{equation}
t=0,
\label{t-solution}
\end{equation}
which implies that the tachyon $t$ is a non-propagating mode
in the AdS$_3 \times$S$^3$ background.
On the other hand we find ${\cal F}_+={\cal F}_-$.
Plugging this with $t=0$ into (\ref{eq-t-ptr}) 
leads to the fact that (\ref{eq-t-ptr}) is trivially satisfied.
Now let us consider the simplest case where all but the tachyon 
$t$ vanish.
In this case, we also find $t=0$ from (\ref{Kalb-Ramond}).
The non-propagation of the tachyon originates from 
the coupling of $f_\pm(T) F_{\pm3}^2$ in (\ref{string-action}) and
the background symmetry (\ref{bck-sol}) of AdS$_3 \times$S$^3$.
Using (\ref{rel-F}) and (\ref{rel-Ftheta}), 
Eqs.(\ref{eq-Phi-ptr}) and (\ref{eq-dilaton-ptr}) lead to 
\begin{eqnarray}
&&\bar \nabla^2 (8\phi-h)
- {4 \over R^2 } h^{i}_{~i}=0,
\label{Phi-ptr1} \\
&&\bar \nabla^2 \phi
- {2  \over R^2 } h^i_{~i}=0.
\label{t-ptr1}
\end{eqnarray}
If $h^i_{~i}=0$, then one finds
\begin{equation}
h=8 \phi,~~ \bar \nabla^2 \phi =0.
\label{eq-dilaton3-b}
\end{equation}
However this corresponds to the linearized equation
for a minimally coupled
scalar.  We need to find the linearized equation for the dilaton.
If $h^i_{~i} = a \phi$, then
Eqs.(\ref{Phi-ptr1}) and (\ref{t-ptr1}) lead to the same equation as
\begin{equation}
\bar \nabla^2 \phi
- {2a  \over R^2 } \phi=0~~~~{\rm with}~~h=6 \phi.
\label{eq-dilaton1-b}
\end{equation}
In order to find $a$, we recall the relation in
(\ref{Phi-chi}).
The original relation comes from the compactification scheme
($g_{10} = e^{\phi_6} g_6 + 
e^{2 \chi} dx_i dx^i$, where $\phi_6 = \Phi - 2 \chi$). 
Here we require the 6D dilaton to be non 
propagating($\phi_6=0$).
Then one finds (\ref{Phi-chi}).
This means that the dilaton is related to the scale of T$^4$.
The linearized version of (\ref{Phi-chi}) should be also 
valid because (\ref{Phi-chi}) is an initial constraint.
This takes the form
\begin{equation}
\phi = 2 \delta \chi.
\label{relation4-b}
\end{equation}
And then using $g_{ij}=\bar g_{ij} + h_{ij}$,
$h^i_{~i} = 8 \delta \chi = 4 \phi$.  This implies $a=4$. 
We use this relation to derive the correct form of the 
dilaton equation instead of keeping $h^i_{~i}$ an independent fluctuation.
Hence the final equation for the dilaton in the string frame takes the form
\begin{equation}
\bar \nabla^2 \phi - {8 \over R^2} \phi =0.
\label{eq-dilaton2-b}
\end{equation}
Also this corresponds to the equation for a minimally coupled scalar with $l=2$
on AdS$_3 \times $S$^3$\cite{Lee98PRD104013}.

\section{Discussions}
\label{discussion}
A new feature of type 0B string theory is the presence of the 
tachyon.
It is known that while the Minkowski vacuum is unstable in 
type 0 string theory, 
the near-horizon geometry of AdS$_5 \times $S$^5$ 
should be a stable background for sufficiently small
radius\cite{Kle99JHEP03015}.
The RR fields work to stabilize the tachyon in the near-horizon.
It is clear from the D5$_\pm$-D1$_\pm$ brane black hole\cite{Lee9907024}
that the near-horizon of AdS$_3 \times$S$^3 \times$T$^4$ is stable 
because $V_\nu(r)$ and $V_t(r)$ take the shapes of the potential barrier.
If there do not exist the RR fields
(${\cal F}_\pm, {\cal H}_\pm$), one finds a potential well for 
the tachyon, which induces an instability in the 
near-horizon\cite{Cha83}.

On the other hand, we find that the tachyon cannot propagate on 
global AdS$_3 \times$S$^3$.
This background corresponds to a non-dilatonic solution.
This means that the 10D dilaton($\Phi$) plays no role in setting 
this background.
However, the propagation of $\Phi$ is alive as shown in 
Eq.(\ref{eq-dilaton2-b}) and its absorption cross section 
was obtained in \cite{Lee9805050,Lee98PRD104013}.
Unfortunately, the tachyon is not a propagating mode.
This comes from the Kalb-Ramond equation, the gauge condition and 
Bianchi identities.
In the s-wave($l=0$) calculation the propagation of tachyon is not allowed.
This contrasts to our naive expectation such that the tachyon is a 
propagating mode in the type 0B string theory.

How we do interpret this non-propagation?
This mainly due to the spacetime symmetry of 
global AdS$_3 \times$S$^3$ background.
We remind the reader that the AdS$_3 \times$S$^3$ background in 
type IIB theory is maximally supersymmetric\cite{Boe98NPB139}.
Although the type 0B theory is not supersymmetric in spacetime, 
the AdS$_3 \times$S$^3$ has the global 
SL(2,{\bf R})$_{\rm L} \times$SL(2,{\bf R})$_{\rm R}\times$SU(2)$_{\rm L}\times$SU(2)$_{\rm R}$ 
group of isometries.
These are part of an AdS supergroup
G=G$_{\rm L}\times$G$_{\rm R}$, where both G$_{\rm L}$ and 
G$_{\rm R}$ contain SU(2)$\times$SL(2,{\bf R}).
This background is indistinguishable from their type IIB 
cousins\cite{Mal98JHEP12005}. 
It seems that these global symmetries prevent the tachyon from 
propagating into the global AdS$_3 \times$S$^3$.
This is clear if the global AdS$_3 \times$S$^3$
(the non-dilatonic solution(\ref{metric})) is compared 
with the D5$_\pm$-D1$_\pm$ brane black hole(the dilatonic solution).
The former has AdS$_3 \times$S$^3$ with asymptotically AdS$_3$ space, 
while the latter takes AdS$_3 \times$S$^3$ only in the 
near-horizon but with asymptotically flat space.
This means that the global AdS$_3 \times$S$^3$ is regarded as the large 
$R$ limit of the near-horizon AdS$_3 \times$S$^3$.
This global spacetime can be achieved only from the special setting as 
(\ref{bck-sol})-(\ref{bck-R}).
As a result the background isometries of global AdS$_3 \times$S$^3$
may protect the tachyon from propagating on this space.

In conclusion, although the tachyon is propagating and stable 
in the near-horizon geometry(AdS$_3\times$S$^3$) of the 
D5$_\pm$-D1$_\pm$ brane black hole,
it is not a propagating mode on the global AdS$_3\times$S$^3$ background.

\section*{Acknowledgement}
This work was supported by the Brain Korea 21
Program, Ministry of Education, Project No D-0025.

\end{document}